\documentclass[12pt,a4paper,oneside]{article}
\usepackage{graphicx,amssymb,amsmath,color}
\usepackage[linkcolor={blue},citecolor={red},colorlinks=true]{hyperref}
\usepackage{enumerate}
\usepackage{ulem}
\usepackage[margin=2cm]{geometry}
\usepackage{slashed,multirow}
\usepackage[utf8]{inputenc}
\usepackage{feynmf}
\usepackage[compat=1.1.0]{tikz-feynman}
\usepackage{hyperref}
\def\ol{\overline}

\def\beq{\begin{equation}}
\def\eeq{\end{equation}}
\def\bea{\begin{eqnarray}}
\def\eea{\end{eqnarray}}

\def\bit{\begin{itemize}}
\def\eit{\end{itemize}}

\def\l{\left}
\def\r{\right}

\def\baa{\begin{array}}
\def\eaa{\end{array}}
\def\ol#1{\overline{#1}}

\def\sl#1{\mathord{\not\mathrel{{\mathrel{#1}}}}}
\def\d{\partial}

\def\simgt{\mathrel{\lower2.5pt\vbox{\lineskip=0pt\baselineskip=0pt
           \hbox{$>$}\hbox{$\sim$}}}}
\def\simlt{\mathrel{\lower2.5pt\vbox{\lineskip=0pt\baselineskip=0pt
           \hbox{$<$}\hbox{$\sim$}}}}

\def\bfc{\begin{figure}\begin{center}}
\def\efc{\end{center}\end{figure}}
\def\nn{\nonumber\\}

\begin{document}

\begin{flushright}
\hspace{3cm} 
SISSA   21/2021/FISI
\end{flushright}
\vspace{.6cm}
\begin{center}

\hspace{1 cm}{\Large \bf  Four-fermion operators at dimension 6: dispersion relations and UV completions
}\\[0.5cm]

\vspace{1cm}{Aleksandr Azatov$^{a,b,c,1}$,  Diptimoy Ghosh$^{d,2}$,  Amartya Harsh Singh$^{d,3}$}
\\[7mm]
 {\it \small

$^a$ SISSA International School for Advanced Studies, Via Bonomea 265, 34136, Trieste, Italy\\[0.15cm]
$^b$ INFN - Sezione di Trieste, Via Bonomea 265, 34136, Trieste, Italy\\[0.1cm]
$^c$ IFPU, Institute for Fundamental Physics of the Universe, Via Beirut 2, 34014 Trieste, Italy\\[0.1cm]
$^d$ Department of Physics, Indian Institute of Science Education and Research Pune, India
 }

\end{center}

\bigskip \bigskip \bigskip

\centerline{\bf Abstract} 
\begin{quote}
A major task in phenomenology today is constraining the parameter space of SMEFT and constructing models of 
fundamental physics that the SM derives from. To this 
effect, we report an exhaustive list of sum rules for 
4-fermion operators of dimension 6, connecting low energy 
Wilson coefficients to cross sections in the UV. Unlike 
their dimension 8 counterparts which are amenable to a 
positivity bound, the discussion here is more involved due to the weaker convergence and indefinite signs  of the dispersion integrals. We illustrate this  by providing examples
with weakly coupled UV completions leading to opposite signs of the Wilson coefficients for both convergent and non convergent dispersion integrals.
We further  decompose dispersion integrals  under 
weak isospin and color groups which lead to a tighter relation between IR measurements and UV models.
These sum 
rules  can become an effective tool for constructing consistent UV 
completions for SMEFT following the prospective 
measurement of these Wilson coefficients.
\end{quote}

\vfill
\noindent\line(1,0){188}
{\scriptsize{ \\ E-mail:
\texttt{$^1$\href{mailto:aleksandr.azatov@NOSPAMsissa.it}{aleksandr.azatov@sissa.it}}, $^2$\href{mailto:diptimoy.ghosh@iiserpune.ac.in}{diptimoy.ghosh@iiserpune.ac.in},
$^3$\href{mailto:amartya.harshsingh@students.iiserpune.ac.in}{amartya.harshsingh@students.iiserpune.ac.in}}
}
\newpage

\tableofcontents

\section{Introduction}

esting the Standard Model(SM) and searching for new physics are two essential goals of the 
current and future  experimental programs in particle physics.  In this respect, all of the 
measurements can be classified as low energy (SM scale ) and high energy experiments.
For low energy observables,  the Standard Model Effective Field Theory( SMEFT) provides an 
excellent tool to consistently parameterize new physical perturbations, classified order by order in the form of non-renormalizable operators with higher dimensions. We expect new physics to kick in above at least the weak scale, and as we approach the regime of high energies greater than this scale, the applicability of EFT techniques becomes successively questionable. Reliable calculations then require a discussion  of the 
explicit  UV completions, and thus it's clear that the connection between UV and IR observables and 
predictions becomes somewhat model dependent, and explicit matching is required to infer useful information. In this direction, dispersion relations provide a model independent way to connect low and high energy 
measurements, in the form of \textit{sum rules} for 
low energy Wilson coefficients and high energy cross sections. This provides a consistent way to match 
the known and measurable low energy, and speculative high energy quantities(for a recent reappraisal see \cite{Adams:2006sv} and for a textbook introduction \cite{Eden:1966dnq,Gribov:2009zz} ). Their power lies in their generality -they follow from the simple and sacred physical requirements of Poincare invariance, unitary and locality.
Recently there has been significant attention directed toward the application of the dispersion relations and sum rules for SMEFT \cite{Falkowski:2012vh,Trott:2020ebl,Gu:2020thj,Zhang:2020jyn}.
For the four fermion interactions  most of the effort so far  has been focused on the dimension 8 operators(\cite{Remmen:2020vts,Bonnefoy:2020yee,Fuks:2020ujk,Gu:2020ldn}) where the sum rules lead to positivity constraints on the Wilson coefficients in a model independent way.

On the other hand, from a phenomenological  point of view, dimension eight operators are very hard to measure at experiments; and most likely the new physics will demonstrate itself first via dimension six corrections to the  SM. Thus, it becomes crucial to understand similar dispersion relations for the dimension six operators.
The situation here is 
drastically different from the dimension 8 discussion because the relevant dispersion integral, aside from being possibly non-convergent, is of indefinite sign and doesn't 
admit any simple model independent positivity bound. 
However, the situation is far from hopeless, and the dispersion relations turn out to be instructive in a different way:
instead of being viewed as a 
constraint on Wilson coefficients, these sum rules are to 
be used as a \textit{tool to constrain the UV completions} of these operators, given signs  to be measured in 
the IR. Therefore, in a way, we are approaching the IR-UV 
relationship from the opposite standpoint to what is 
customary. Our emphasis is on model building for a full 
theory by taking IR measurements as our input, instead of 
trying to predict these measurements from general inputs 
from the UV theory.
We will show that different signs of the Wilson 
coefficients  will be related to the dominance of the 
particle collision cross sections in the various channels, and decompose these cross sections as explicitly as 
possible to indicate the quantum numbers of initial states with dominant cross sections. Moreover, it is crucial to 
emphasize that sum rules can only be written down for a 
subspace of the dimension 6 basis, namely the effective 4 
fermion operators that can generate forward amplitudes. 
Based on these sum rules, we will report  examples of the 
weakly coupled UV completions which can lead to 
either sign of the Wilson coefficients. Such information, 
which we believe was not consistently summarized before, 
can become a useful guide  for the future measurements in 
case some of the Wilson coefficients are discovered to be 
non zero. These measurements, supplemented with the sum 
rules we derive, will 
bring us closer to an 
understanding of the fundamental physics which the SMEFT 
derives from. 

The manuscript is organized as follows : in section \ref{sec:disp} we briefly review dispersion integrals. In section \ref{sec:eeee}, we study in detail the operator $(\bar{e}_R\gamma^\mu e_R)^2$ and illustrate the relation between UV completions and signs of the effective operator at tree level and 1 loop. In section \ref{sec:4f}, we present the whole set of the four fermion operators and identify which of them can be constrained by the dispersion relations. 
Results are summarized in the section \ref{sec:sum}.
Most details of the calculations have been relegated to the appendices. 

\section{Review of dispersion relations}\label{sec:disp}

In this section, we will review dispersion relations and their applications to constraints on EFTs following the discussion in \cite{Adams:2006sv,Low:2009di,Falkowski:2012vh,Bellazzini:2014waa}( readers familiar with the formalism can proceed directly to section \ref{sec:eeee}).
It is a general principle that the non-analyticities associated with scattering amplitudes have a physical origin in the form of poles and branch cuts arising from localized particle states, and thresholds. The positivity of the spectral function in the Kallen-Lehmann decomposition generalizes to more general cross sections, which can be related to elastic forward scattering amplitudes via a dispersion integral, to be reviewed in a moment. What this means in an EFT context is that, \textit{in perturbation theory}, one can evaluate the 2 sides of a dispersion integral to a certain order; allowing us to extract information about the effective IR coupling that contributes to that amplitude at low energies on one side of the relation, from general observations about the UV piece of the dispersion relation without any explicit matching. 

While unitarity reflects in the positivity of the spectral function and cross sections, we need additional information about the high 
energy behaviour of the amplitude to control the 
dispersion integral at the infinite contour. The asymptotics of amplitudes at high energies is a question about the unitarity and locality of the theory. The 
famous Froissart bound-whilst technically proved only for 
theories with a mass gap, but believed to hold true 
generally- tells us that the behaviour of the amplitude 
$A(s)$ is such that $A(s)/s^2\to 0$ as $s\to \infty$ 
(
\cite{Froissart:1961ux,Martin:1962rt,Martin:1965jj}
).  This, in general allows us to 
write down a dispersion relation with 2 subtractions, i.e. a linear polynomial of the form $a(t)+b(t)s$ supplemented 
by a contour integral picking up the nonanalytic structure of the amplitude. $a(t),b(t)$ cannot be determined by unitarity 
alone, but the nonanalytic structure can be related to 
manifestly positive cross sections via the optical 
theorem. We can then differentiate this relation w.r.t $s$ twice to get rid of the unknown subtractions, and we're 
left with a manifestly positive integral on the right, and the coeffecient of $s^2$ in $A(s)$ on the left-therby 
leading to what are conventionally called 'positivity 
bounds' \cite{Adams:2006sv} 
on EFT parameters. 

This prescription, however, cannot be directly applied to dimension $6$ operators.
Their contribution to $2\to 2$ amplitudes scales as $p^2$, and so $d^2A(s)/ds^2$ 
kills information about their couplings, and we cannot constrain them in any way. The best we can do is to look at $dA(0)/ds$, and be left with a dispersion integral {of indefinite sign}
as well as an undetermined subtraction constant (which we'll call $C_\infty$, as it captures the pole 
of the amplitude at infinity).  
 
\begin{figure}[htp]
    \centering
    \includegraphics[scale=1]{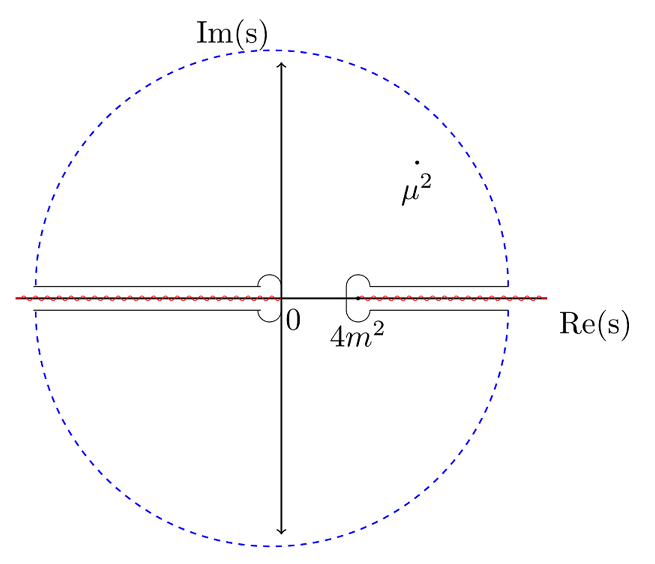}
    \caption{Analytic structure in complex $s$ plane. The infinite circle is centered at $2m^2$, and will be traversed counterclockwise.\label{fig:cont}
    }
    \end{figure}

Let us briefly derive this dispersion relation 
from first principles. Consider a process $ab\to ab$ with the amplitude $A_{ab\to ab}\equiv A_{ab}(s,t)$, and in the forward limit ($t\to0 $). This amplitude can be expanded as
\begin{equation}
A_{ab}(s,0)=\sum_n c_n(\mu^2) (s-\mu^2)^n,~~~{c_n(\mu^2)=\frac{1}{n!}\frac{\d^n}{\d s^n}A_{ab}(s,0)|_{s=\mu^2}}
\end{equation}
 about some arbitrary reference scale $\mu^2$ where the amplitude is analytic. We can now use Cauchy's theorem to write

\bea
\label{eq:dispbasic}
\frac{1}{2\pi i}
\oint ds\frac{A_{ab}(s,0)}{(s-\mu^2)^{n+1}} =\sum_{s_i,\mu^2} Res \frac{A_{ab}(s,0)}{(s-\mu^2)^{n+1}}
=c_n(\mu^2)+\sum_{s_i} Res \frac{A_{ab}(s,0)}{(s-\mu^2)^{n+1}},
\eea

where $s_i$ are the 
physical poles associated with IR stable resonance 
exchanges in the scattering, and the contour of 
integration is shown on the Fig. \ref{fig:cont}.
The residues at physical poles are IR structures that we will drop henceforth.
This can always be done if the scale $\mu$ is chosen such that $\mu^2\gg m_{IR}^2$, where $m_{IR}^2$ corresponds to the scale of the $s_i$ poles. Indeed, the last term in Eq.\ref{eq:dispbasic} gives corrections of the order $\mathcal{O}(m_{IR}^2/\mu^2)$, which can be safely ignored.

The analytic structure of the amplitude allows to decompose the integral as a sum of the contributions along the branch cuts and over infinite circle, so that scehamtically
\bea
&&\frac{1}{2\pi i}\int ds\frac{A_{ab}(s,0)}{(s-\mu^2)^{n+1}}=\textrm{integrals along cuts}+\textrm{integral on big circle}=C^n_\infty +I_n\nn
&&C^n_\infty=\int^{2\pi}_0 \frac{d\theta}{2\pi}\frac{A_{ab}(|s_\Lambda|e^{i\theta},0)}{(|s_\Lambda|e^{i\theta}-\mu^2)^{n+1}}\cdot(|s_\Lambda|e^{i\theta})
\eea
The integration over the branch cuts can be written as a sum of the integrals over discontinuities
\begin{equation}
2\pi iI_n=\int_{4m^2}^\infty\bigg( \frac{A_{ab}(s+i\epsilon,0)-A_{ab}(s-i\epsilon,0)}{(s-\mu^2)^{n+1}}+(-1)^n\frac{A_{ab}(4m^2-s-i\epsilon,0)-A_{ab}(4m^2-s+i\epsilon,0)}{(s-4m^2+\mu^2)^{n+1}}\bigg).
\end{equation}
Since $4m^2-s=u$ for $t=0$, the second term is just the $u$ channel crossed amplitude for the process $a\bar{b}\to a\bar{b}$ i.e. $A_{a\bar{b}}$(instead of $ab\to ab$)\footnote{ 
Crossing relations for particles with spin become more nontrivial  (see for example \cite{Bellazzini:2016xrt,deRham:2017zjm}). However, in the case of the massless spin $1/2$ particles, which are the interest of this paper, the usual crossing relations for the forward amplitude remain valid \cite{Bellazzini:2016xrt} and we will not worry about these issues in the rest of the paper.}.
Using the optical theorem, we can  rewrite the discontinuity in terms of cross section 
and in the limit    $m\to 0$, and $\mu\to 0$
we obtain :
\begin{equation}
\label{eq:dispN}
I_n=\int\frac{ds}{\pi s^{n}}\bigg(\sigma_{ab}+(-1)^n\sigma_{a\bar{b}}\bigg).
\end{equation}

For dimension six operators, we will be interested in dispersion relations of Eq.
\ref{eq:dispbasic} for the case $n=1$.
\bea
c_1(\mu^2)
=\int \frac{ds}{\pi s}\bigg(\sigma_{ab}-\sigma_{a\bar{b}}\bigg)+C_\infty^{(n=1)}.
\eea
Note that the quantity $c_n(\mu^2)$ on the left hand side can be evaluated in IR  using the EFT expansion.
This introduces an additional source of corrections of the order ${\cal O}(\mu^2/\Lambda^2)$, where $\Lambda$ is the scale suppressing higher dimensional operators. 
We can see that the dispersion relations are valid up to corrections of the order ${\cal O}(m_{IR}^2/\mu^2 , \mu^2/\Lambda^2)$, and these can be ignored if
 $\Lambda^2\gg\mu^2\gg m^2_{IR}$.

At last, let us mention that the forward limit $t\to 0$ 
must be taken with care, and is in principle problematic 
in the presence of massless particles propagating in the 
t-channel of the UV amplitude (see for example \cite{Falkowski:2012vh,Bellazzini:2014waa}). 
In fact, we always have the usual SM Coulumb singularities that lead to the bad behaviour in the forward limit. The way out of this problem is by using IR mass regulators to match the known SM contributions to both sides of the dispersion relation, and subtract them away.


\section{Warm up exercise}
\label{sec:eeee}
We consider the simplest case of a fully right handed operator which is made up of singlet fields $e_R$, all of the same generation (the dispersion relation for this operator was presented in  \cite{Gu:2020thj} too),  
\bea
\label{eq:eeee}
c_{RR}(\bar e_R\gamma_\mu e_R)(\bar e_R \gamma^\mu e_R).
\eea
Following  the strategy outlined in the previous section, we start by considering the amplitude  $A_{e \bar e}$ and derive the following dispersion relation
\begin{equation}
\frac{dA_{e_R \ol {e_R}}(s,0)}{ds}\bigg|_{s=0}=
\int\frac{ds}{\pi s}\l(\sigma_{e_R \ol{e_R}}-\sigma_{e_R e_R}\r)+C_\infty ,
\end{equation}
where we have omitted the $(n=1)$ subscript for $C_\infty$.
The amplitude in the IR ($s\to 0$) limit can be safely calculated using the EFT and we find (we use helicity amplitudes; for notations and for the explicit conventions see appendix \ref{sec:aphelicity}):
\bea
&&A_{e_R \ol{e_R}}(s,t)=c_{RR}\cdot2([2\gamma_\mu 1\rangle[3\gamma^\mu4 \rangle-[3\gamma_\mu1\rangle[2\gamma^\mu 4\rangle)=-8c_{RR}[23]\langle 14\rangle\nonumber\\
&&A_{e_R \ol{e_R}}(s,t)|_{t\to 0}=-8 c_{RR}s
\eea
so that we arrive at the following sum rule for the $c_{RR}$ Wilson coefficient\footnote{In this expression we should take the value of the Wilson coefficient at the scale $\mu\to 0$. The RGE evolution of the Wilson coefficients from the EFT cut off scale to $\mu$ can lead to the modification of the Eq.\ref{eq:dispmaster} (see \cite{Chala:2021wpj} for a recent discussion). In this paper we will assume that these running effects are subleading and can be safely ignored.}
\begin{equation}
\label{eq:dispmaster}
-8c_{RR}=\int\frac{ds}{\pi s}\l(\sigma_{e_R \ol{e_R}}-\sigma_{e_R e_R}\r)+C_\infty
\end{equation}
Let us see how this equation can be used as guidance for UV completions that lead to the possible signs of the $c_{RR}$ Wilson coefficient. 

\subsection{Charge neutral vector exchange}\label{sec:Z}
Let us start with the  negative sign for $c_{RR}$. The dispersion relation predicts that this will be generated by the models with resonances in $e \bar e$ channel (apart from the $C_\infty$ contribution). The simplest model which can enhance the $\sigma_{e \bar e}$ cross section is a simple $Z'$ with the interaction
\bea
{\cal L}_{Z'}=\lambda Z'_\mu\bar{e}_R\gamma^\mu e_R.
\eea
Integrating $Z'$ at tree level we obtain for the Wilson coefficient 
\bea
c_{RR}=-\frac{\lambda^2}{2M_{Z'}^2}\label{eq:crr},
\eea
where the sign follows the prediction of the dispersion relations. 
However, inspecting the amplitudes carefully, we see that the massive vector exchange in the $t-$ channel spoils the convergence of the amplitude in the forward region, making the integral over the infinite circle non-vanishing.
To this end, let us look at the amplitude $A_{\bar{e}_R e_R}$ in detail- 
\bea
iA=-\lambda^2\bigg([2\gamma^\mu 1\rangle[3\gamma_\mu 4\rangle\frac{-i}{s-M_{Z'}^2}-[3\gamma^\mu 1\rangle[2\gamma_\mu 4\rangle\frac{-i}{t-M_{Z'}^2}\bigg)\nonumber\\
A(s,t)=-2\lambda^2[23]\langle 14\rangle\bigg(\frac{1}{s-M_{Z'}^2}+\frac{1}{t-M_{Z'}^2}\bigg).
\eea
In the forward limit, this amplitude goes as 
\bea
A(s,t)|_{t\to 0}=-2 \lambda^2 s\l(\frac{1}{s-M_{Z'}^2}+\frac{1}{-M_{Z'}^2}\r).\label{eq:AmpZ}
\eea
We can see that the integral over infinite contour becomes non zero and is equal to 
\bea
C_\infty^{(Z')}=\frac{2 \lambda^2}{ M_{Z'}^2}.\label{eq:pole}
\eea
We see that even though the 
contribution from the infinite contour is
non-zero, it turns out of the same sign and size as the cross section part of the dispersion relation 
\bea
\l[\int\frac{ds}{\pi s}\l(\sigma_{e_R \ol{e_R}}-\sigma_{e_R e_R}\r)\r]^{(Z')}=\frac{2\lambda^2}{M_{Z'}^2},
\eea
(see appendix \ref{sec:brutef} for details of the calculation). The fact that exchange of the elementary vector boson spoils the convergence of the amplitude in the forward limit at large $s$ is not new and was observed  for example in  \cite{Falkowski:2012vh} in the discussion of  the other dimension six operators. 

Let us extend the discussion for the operators with two fermion flavours. For example $c_{e\mu}(\bar{e}_R\gamma^\mu e_R)(\bar{\mu}_R\gamma_\mu \mu_R)$  contributes  $e\bar{\mu}\to e\bar \mu$ in the IR. This operator can be generated by two kinds of  UV completions with a charge neutral vector boson-
\begin{equation}
    \mathcal{L}_{UV}^{(1)}=\lambda Z_{(1)}^\mu(\bar{e_R}\gamma^\mu \mu_R+h.c)\hspace{7mm}\mathcal{L}^{(2)}_{UV}=(\lambda_1Z_{(2)}^\mu\bar{e}_R\gamma^\mu e_R+\lambda_2Z_{(2)}^\mu \bar{\mu}_R\gamma_\mu \mu_R)\label{eq:ij}
\end{equation}
The analysis in both cases is very similar to the single flavour discussion; however, in the first case (${\cal L}_{UV}^{(1)}$) the integral over infinite contour vanishes, since there is no amplitude with $Z_{(1)}$ in the $t$-channel. Writing down the dispersion relations for the $e \bar \mu\to e\bar \mu$ scattering we will obtain (note that there is a different numerical prefactor compared to Eq.\ref{eq:dispmaster} due to combinatorics):
\bea
c_{e\mu}=-\frac{1}{2}\l[\int \frac{ds}{\pi s}\l(\sigma_{e_R \bar \mu_R}-\sigma_{e_R \mu_R}\r)\r]=-\frac{|\lambda |^2}{M_{(1)}^2}.
\eea
In the second case $({\cal L}_{UV}^{(2)})$, we are in the opposite situation since both cross sections $\sigma_{e\bar{\mu} (\mu)}=0$ vanish at leading order
in perturbation theory.
However there is  a forward amplitude for this process, which comes from   $t$-channel diagram and it contributes only to $C_\infty$. In other words, \textit{the pole at infinity saturates the dispersion relation, and even though no corresponding UV cross section can be measured to constrain this coefficient, it can be nonzero} because of this pole. In fact, a simple calculation yields 
\begin{equation}
\label{eq:zprime2}
   c_{e\mu}=-\frac{C_\infty}{2}=-\frac{\lambda_1\lambda_2}{M_{(2)}^2}
\end{equation}
which can be either positive or negative depending on the values of the $\lambda_1,\lambda_2 $ couplings. 
Let us continue with our examination of the UV completions for the various signs of the $c_{RR}$

\subsection{Charge two scalar}\label{sec:scalar}
What about the positive sign of $c_{RR}$? The dispersion relation in Eq. \ref{eq:dispmaster} predicts  that this happens for  UV completions  that generate only $\sigma_{ee}$ cross section.
The simplest possibility is a charge two scalar with the interaction
\bea
{\cal L}=\kappa \phi \ol{e_R^c} e_R+ h.c.
\eea
Then at the order ${\cal O}(\kappa^2)$,  only $\sigma_{ee}$ will be non-vanishing, so the Wilson coefficient must be positive. Indeed, integrating out the scalar field at tree level gives
\bea
c_{RR}=\frac{|\kappa|^2}{2M_\phi^2}\label{eq:crr'}
\eea
 which is manifestly positive. In this case the forward amplitude converges quickly enough, so that $C_\infty=0$  -this is just the statement 
that a scalar cannot be 
exchanged in the $t$ channel 
when opposite helicity fermions are scattered.
We see that the both signs of the Wilson coefficient are possible with a weakly coupled UV completion.
One can still wonder whether the negative sign of the $c_{RR}$ interactions in the Eq. \ref{eq:crr} is related to the $t-$ channel pole and non-convergence of the amplitude in the UV. To quell any doubts,  in the next subsection we will build  a weakly coupled UV completion without new vector bosons and with  convergent forward amplitudes.

\subsection{UV completion at 1-loop}\label{sec:loop}
Let us extend the SM with vector-like fermion $\Psi$ of charge $1$ and a charge $-2$ complex scalar $\phi$ with a Yukawa interaction
\bea
{\cal L}= |D\phi|^2+i \bar \Psi \sl D \Psi+ M_\psi \bar \Psi \Psi- M^2_\phi |\phi|^2+ y \bar e_R \phi \Psi .
\eea
This generates an effective operator at the order $O(y^4)$, and at this order the only cross section available is $\sigma_{e \bar e}$. The dispersion relation predicts that the Wilson coefficient  must be negative. Moreover, $C_\infty=0$ here as the amplitude scales slowly enough with $s$. Indeed, integrating out heavy fields at one loop
we obtain 
\bea
&&c_{RR}=-\frac{|y|^4}{128 \pi^2 M_\Psi M_\phi}f(x),~~~x\equiv \frac{M_\Psi}{M_\Phi}\nn &&f(x)=\frac{(x+4x^3\log x-x^5)}{(1-x^2)^3},~~~~~\lim_{x\to 1}f(x)=1/3\label{eq:cbox}
\eea
where one can see that the function $f(x)$ is always positive.
See appendix \ref{sec:brutef} for explicit verification of the dispersion integral in the case $M_\Psi=M_\Phi$.

In summary, this warm up exercise shows us that both signs of the Wilson coefficients are possible within weakly coupled theories.  Contribution of the infinite contours is important for the 
t-channel exchange of the vector 
resonances. Interestingly, both signs of 
the Wilson coefficient are possible even 
for the weakly coupled models with 
vanishing $C_\infty$
\footnote{This result for the Wilson coefficients  contradicts  the findings of the Ref.\cite{Remmen:2020uze}, where the only possible sign of the Wilson coefficient was found to be positive.}
In the following, 
we will derive the set of the dispersion 
relations for the whole set of four 
fermion operators and identify the UV 
completions leading to the various signs 
of the Wilson coefficients. 

\section{Four fermion operators}\label{sec:4f}
First of all, let us define a complete basis of the 
 four fermion operators, and we will do this following  the notations of the Ref. \cite{Grzadkowski:2010es}, \cite{deBlas:2017xtg}: 
\bea
\label{eq:basis1}
&&\hbox{\bf purely left-handed}\nn
 &&   O_{ll}^{ijkm}  =     \left(\bar{l^i}_L \gamma_\mu l^j_{L}\right)
      \left(\bar{l^k}_L \gamma^\mu l^m_L\right),~~~
      O_{qq}^{(1)ijkm}=\left(\bar{q^i}_L \gamma_\mu q^j_L\right)
      \left(\bar{q^k}_L \gamma^\mu q^m_L\right),\nn
&&  O_{qq}^{(3)ijkm}=\left(\bar{q^i}_L \gamma_\mu \sigma_a q^j_L\right) \left(\bar{q^k}_L \gamma^\mu \sigma_a q^m_L\right),~~
O_{ql}^{(1)ijkm}    =  \left(\bar{l^i}_L\gamma_\mu l^j_L\right)
      \left(\bar{q^k}_L \gamma^\mu q^m_L\right)\nn
 &&     O_{ql}^{(3)ijkm}    =  \left(\bar{l^i}_L\gamma_\mu \sigma_a l^j_L\right)
      \left(\bar{q^k}_L \gamma^\mu \sigma_a q^m_L\right),\nn
 &&     \hbox{\bf purely right-handed}\nn
 &&     O_{ee}^{ijkm}=\left(\bar{e}_R \gamma_\mu e_R\right)
      \left(\bar{e}_R \gamma^\mu e_R\right),~~~
      O_{uu}^{ijkm    
      }=\left(\bar{u}_R \gamma_\mu u_R\right)
      \left(\bar{u}_R \gamma^\mu 
      u_R\right)\nn
&&O_{dd}=      \left(\bar{d}_R \gamma_\mu d_R\right)
      \left(\bar{d}_R \gamma^\mu d_R\right),~~
O_{ud}=\left(\bar{u}_R \gamma_\mu u_R\right)
      \left(\bar{d}_R \gamma^\mu d_R\right)\nn
&& O_{ud}^{(8)} =  \left(\bar{u}_R \gamma_\mu T_A u_R\right)
      \left(\bar{d}_R \gamma^\mu T_A d_R\right),~~~
O_{eu}=\left(\bar{e}_R \gamma_\mu e_R\right)
      \left(\bar{u}_R \gamma^\mu u_R\right)\nn
 &&   O_{ed}  \left(\bar{e}_R \gamma_\mu e_R\right)
      \left(\bar{d}_R \gamma^\mu d_R\right),
      \eea
\bea
\label{eq:basis2}
&&\hbox{\bf left-right}\nn
&& O_{le}= \left(\bar{l}_L \gamma_\mu l_L\right)
      \left(\bar{e}_R \gamma^\mu e_R\right),~~~
    O_{qqee}  \left(\bar{q}_L \gamma_\mu q_L\right)
      \left(\bar{e}_R \gamma^\mu e_R\right)
      \hspace{3.2 cm}
      \nn
&&      
 O_{lu}=     \left(\bar{l}_L \gamma_\mu l_L\right)
      \left(\bar{u}_R \gamma^\mu u_R\right),~~~
O_{ld}=\left(\bar{l}_L \gamma_\mu l_L\right)
      \left(\bar{d}_R \gamma^\mu d_R\right)\nn
      &&
 O_{qu}^{(1)}=     \left(\bar{q}_L \gamma_\mu q_L\right)
      \left(\bar{u}_R \gamma^\mu u_R\right),~~~
 O_{qu}^{(8)}=\left(\bar{q}_L \gamma_\mu T_A q_L\right)
      \left(\bar{u}_R \gamma^\mu T_A u_R\right)\nn
 &&
   O_{qd}^{(1)}=  \left(\bar{q}_L \gamma_\mu q_L\right)
      \left(\bar{d}_R \gamma^\mu d_R\right),~~~
 O_{qd}^{(8)} =   \left(\bar{q}_L \gamma_\mu T_A q_L\right)
      \left(\bar{d}_R \gamma^\mu T_A d_R\right)\nn
   &&  O_{ledq}= \left(\bar{l}_L e_R\right)
      \left(\bar{d}_R q_L\right),~~~
      O_{quqd}^{(1)}=\left(\bar{q}_L  u_R\right) i\sigma_2
      \left(\bar{q}_L d_R\right)^{\mathrm{T}}   \nn   
   &&
   O_{lequ}^{(1)}= \left(\bar{l}_L  e_R\right) i\sigma_2
      \left(\bar{q}_L u_R\right)^{\mathrm{T}},~~
O_{lequ}^{(3)}= \left(\bar{l}_L  \sigma_{\mu\nu} e_R\right) i\sigma_2
      \left(\bar{q}_L \sigma^{\mu\nu} u_R\right)^{\mathrm{T}}\nn
   &&    O_{quqd}^{(8)}=\left(\bar{q}_L T_A  u_R\right) i\sigma_2
      \left(\bar{q}_L  T_A d_R\right)^{\mathrm{T}}, 
\eea
\bea
\label{eq:basis3}
&&\hbox{\bf baryon number violating}\nn
  &&  O_{duq}=    \epsilon_{ABC} \left(\bar{d}^{c\,A}_R u^B_R\right)
        \left(\bar{q}^{c\,C}_L i\sigma_2 l_L\right),~~
       O_{qqu}= \epsilon_{ABC} \left(\bar{q}^{c\,A}_L i\sigma_2 q^B_L\right)
        \left(\bar{u}^{c\,C}_R e_R\right)\nn
&&    O_{duu}=    \epsilon_{ABC} \left(\bar{d}^{c\,A}_R u^B_R\right)
        \left(\bar{u}^{c\,C}_R e_R\right),~~
O_{qqq} \epsilon_{ABC} (i\sigma_2)_{\alpha\delta} (i\sigma_2)_{\beta\gamma}
      \left(\bar{q}^{c\,A\alpha}_L q^{B\beta}_L\right)
      \left(\bar{q}^{c\,C\gamma}_L l^{\delta}_L\right).
\eea
The rest of the possible operators can be reduced via some Fierzing to the
basis of Eq.\ref{eq:basis1}-\ref{eq:basis2}-\ref{eq:basis3}, using the completeness relations for the SU(2) and SU(3) generators
\begin{align}
&\sum_{a=1}^{3}\left(\sigma^{a}\right)_{i j}\left(\sigma^{a}\right)_{k l}=2\left(\delta_{i l} \delta_{k j}-\frac{1}{2} \delta_{i j} \delta_{k l}\right) \\
&\sum_{A=1}^{8}\left(T^{A}\right)_{i j}\left(T^{A}\right)_{k l}=2\left(\delta_{i l} \delta_{k j}-\frac{1}{3} \delta_{i j} \delta_{k l}\right)
\end{align}

As we have seen in the previous section, the dispersion relations are effective in the case of forward scattering i.e. when the initial and final states are the same
\footnote{Recently it was shown that the scattering of the mixed(entangled) flavour states can lead to the additional constraints in the case of the dimension eight operators \cite{Remmen:2020vts,Bonnefoy:2020yee}, where strict positivity bounds can be applied. In the case of dimension six operators the measurements of the cross sections for the mixed states looks almost impossible, so we do not investigate this direction further.}.
Therefore, only the following subspace of operators can be subject to sum rules -
\bea
O_{ll}^{iijj},O_{ll}^{ijji},O_{qq}^{(1,3)iijj},O_{qq}^{(1,3)ijji}, O_{ql}^{(1,3)iikk},O_{ee,uu,dd}^{iijj},O_{ee,uu,dd}^{ijji},\nn O_{ud}^{(1),(8),iijj}, O_{ed}^{iijj}, O_{eu}^{iijj},
O_{le,qe,lu,ld}^{iijj}, O^{(1)(8)iijj}_{qu}, O^{(1)(8)iijj}_{qd},
\eea
which will be the focus of this paper. For fully right-handed operators, the discussion follows closely the results reported for $O_{ee}$ above. Therefore, we will henceforth report only the results and the examples of UV completions leading to various signs. 
\vspace{5mm}

\subsection{Experimental constraints}
Having defined the operators which we will consider in our discussion, let us briefly mention the status of the experimental bounds based on the discussion  in \cite{Falkowski:2015krw,Falkowski:2017pss}. Current bounds on on four lepton  and two lepton two quark operator come  from the combinations of the $Z,W$ pole observables, fermion production at LEP, low energy neutrino scatterings , parity violating electron scatterings, and parity violation in atoms. One of the challenges in deriving these bounds comes from the modifications of $W,Z$ vertices which too can contribute to the same low energy observables, so that the global fit including the $W,Z$ pole observables becomes necessary. For example, for two lepton two quark operators  Ref. \cite{Falkowski:2017pss}
has found nine flat directions unbounded experimentally. Current combinations of the low energy experimental constraints as well as LHC measurements bound the various Wilson coefficients in the range $10^{-2}-10^{-3}$  (where the operators are assumed to be suppressed by the $v_{ew}^2$ scale), which means sensitivity to the scales $\mathcal{O}$(few TeV). Just to be specific, for example the four-electron operator discussed in Eq. \ref{eq:eeee} is bounded by the Bhabha scattering measurements at LEP-2\cite{ALEPH:2013dgf} and SLAC E158 experiment for the M\o{}ller scattering ($e^-e^-\to e^-e^-$)\cite{SLACE158:2005uay}, where both experiments are testing the complementary combinations of the Wilson coefficients leading to the net sensitivity of $\sim 4\times 10^{-3} v_{ew}^{-2}$ on the value of the Wilson coefficient. 
LHC measurements of the dilepton production in $pp$ scattering leads to additional strong constraints on the two quark-two lepton operators \cite{Alioli:2017nzr,Farina:2016rws}, where for some operators we will become sensitive to new physics up to the scale of $\sim 50$ TeV.
So far, all of the measurements are consistent with SM predictions.

\subsection{FULLY RIGHT HANDED}

\subsubsection{$O_{ee}$}
This operator has already been discussed in the section \ref{sec:eeee} and we would just like to emphasize that there are no sum rules for more than two flavours of fermions. Following the notations of Eq. \ref{eq:basis1}-\ref{eq:basis2} the dispersion relations can be summarized as:
\begin{align}
-8c_{ee}^{iiii}=\int \frac{ds}{\pi s}\bigg(\sigma_{\bar{e}_ie_i}-\sigma_{{e}_ie_i}\bigg)+C_\infty\\
 -2(c^{iijj}_{ee})|_{i\neq j}=\int\frac{ds}{\pi s}\bigg(\sigma_{\bar{e}_ie_j}-\sigma _{\bar{e}_i\bar{e}_j}\bigg)+C_\infty \label{eq:ee}
\end{align}
Note that in this simple case where the fields are singlets, the operators $O_{ee}^{iijj}$ and $O_{ee}^{ijji}$ are identical after Fierzing; and $O_{ee}^{iijj}$ and $O_{ee}^{jjii}$ are just trivially identical by symmetrization, so we report the dispersion relation only in terms of $c_{ee}^{iijj}$ in order to not double-count.
Summarising the discussion about UV completions in the section \ref{sec:eeee} we have:
\paragraph{$c_{ee}^{}< 0$}:
neutral $Z'$ at tree level; Vectorlike singlet fermion $\Psi$ and a heavy singlet comlex scalar $\Phi$ with $Q[\Phi \Psi]=-1$ at 1 loop.
\paragraph{$c_{ee}> 0$}:
Charge 2 scalar; 
for different flavours ($O_{ee}^{iijj}|_{i\neq j}$), $Z'$ can lead to a possibly  positive sign as well  if the couplings to the different flavours of the fermions are of opposite signs  (see Eq.\ref{eq:zprime2})


\subsubsection{$O_{uu}, O_{dd}$ }
Let us proceed with our investigation of 
the  four fermion quark operators.
The discussion proceeds exactly in the same way as for the leptons,  except for new color structure.  Fierzing them into the basis of Eq.\ref{eq:basis1}-\ref{eq:basis2} there are only six structures of the operators $O_{uu,dd}^{iijj},O_{uu,dd}^{ijji}$, which are in this case not related by a Fierz identity because of an implicit contraction of color indices. Let us start with the operators where all of the quarks have the same hypercharge, and focus on the operator
$O_{uu}^{iiii}$. Denoting by $\alpha,\beta$ the color indices  and considering same and different color scatterings, we will obtain  the following relations:
\bea
-8 c_{uu}^{iiii}=\int\frac{d s}{\pi s}
\l(\sigma_{u_\alpha \bar u_\alpha}- \sigma_{u_\alpha  u_\alpha}\r)+ C_\infty^{\alpha \alpha}
=\int \frac{ds}{\pi s}\l(\frac{2\sigma_{u \bar u}^{(8)}+\sigma^{(1)}_{u\bar u }}{3}-\sigma_{uu}^{(6)}\r)- C_\infty^{uu,(6)}
\nn
-4 c_{uu}^{iiii}=\l[\int\frac{d s}{\pi s}
\l(\sigma_{u_\alpha \bar u_\beta}- \sigma_{u_\alpha  u_\beta}\r)+ C_\infty^{\alpha \beta}\r]_{\alpha\neq \beta}=\int\frac{ds}{\pi s}\l(\sigma_{u \bar u}^{(8)}-\frac{\sigma_{uu}^{(\bar 3)}+\sigma^{(6)}_{uu}}{2}\r)+ C_\infty^{u\bar u,(8)}.
\eea
In the last step, we have decomposed the various possibilities of the initial state fermions in terms of the $SU(3)$ QCD representations. This is convenient,  
since  the Wigner-Eckart theorem requires the amplitudes to remain the same for all of the components of the irreducible representation. In particular, for the quark antiquark scattering  the initial state will always be decomposed as a singlet and octet  of $SU(3)$.
Even though measuring $\sigma^{(8)}$ and $\sigma^{(1)}$ independently at collider experiment looks practically impossible, such dispersion relations can become very useful for model building  if the non-zero values of the Wilson coefficients are found.
Note   that we  can calculate the integral over  the infinite contour using amplitude $A_{u\bar u}$ or its crossed version $A_{u  u}$ and the values of this integrals will satisfy (see appendix \ref{sec:WE} for details):
\bea
-C_{\infty}^{uu (6)}=\frac{2C_\infty^{u \bar u (8)}}{3}+\frac{C_\infty^{u \bar u(1)}}{3}\nn
-\frac{C_\infty^{uu(6)}+C_\infty^{uu(\bar 3)}}{2}=C_{\infty}^{u \bar u (8)}.
\eea
Re-expressing everything in terms of the color averaged cross section we will obtain
\bea
-\frac{16}{3} c_{uu}^{iiiii}=\int\frac{d s}{\pi s}
\l(\sigma_{u \bar u}- \sigma_{u  u}\r)- \frac{1}{3}C_\infty^{uu(6)} +\frac{2}{3}C_\infty^{u \bar u(8)}
\eea
Again, $C_\infty$ can be non-vanshing, for example, in  UV models with charge neutral vector  resonances exchange in the $t$ channel, but unlike the four electron case here this resonance can be either singlet or octet of $SU(3)$QCD.
Extending this analysis to the case of different flavour of the up quarks  we will obtain :
\bea
\label{eq:uu}
&&-2 c_u^{iijj}=\int \frac{ds}{\pi s}\l(\frac{2\sigma_{u \bar u}^{(8)}+\sigma^{(1)}_{u\bar u }}{3}-\sigma_{uu}^{(6)}\r)- C_\infty^{uu,(6)}
\nn
&&-2 (c_u^{iijj}+c_u^{ijji})=\int\frac{ds}{\pi s}\l(\sigma_{u \bar u}^{(8)}-\frac{\sigma_{uu}^{(\bar 3)}+\sigma^{(6)}_{uu}}{2}\r)+ C_\infty^{u\bar u,(8)},
\eea
We again mention that the operators $O^{iijj}_{uu}$ and $O_{uu}^{jjii}$ (similarly for $O^{ijji}_{uu}$ and $O^{jiij}_{uu}$)are trivially identical, so it's important that we don't double count them.
As before, expressing everything in terms of uncolored cross sections, we find
\bea
-2c^{iijj}_{uu}-\frac{2}{3}c^{ijji}_{uu}=\int\frac{ds}{\pi s}\bigg(\sigma_{u\bar{u}}-\sigma_{uu}\bigg)
+\frac{8}{9}C^{u \bar u(8)}_{\infty}+\frac{1}{9}C^{u \bar u (1)}_{\infty}
\eea
and exactly the same relations hold for the down quarks.

Let us look at the possible UV completions. In the case of $c_{uu}^{iiii}$, we will have a negative sign of the Wilson coefficient with $Z'$, and a positive sign for the charge $-4/3$ scalar. Similar to the lepton case, we can generate a negative Wilson coefficient by adding vectorlike fermions and a complex scalar with $Q[\Phi \Psi]=2/3$ and $(\Phi \psi)-$ fundamental of QCD. The discussion of two fermion flavours is almost identical to the lepton case. 

To demonstrate an explicit verification of these sum rules, in Appendix (\ref{sec:octet}), we provide an example of a UV-completion of the type $g V^A_\mu(\bar{u}_i\gamma^\mu T^A u_i)$. This is a flavor diagonal interaction with a color octet vector and a universal coupling; where the sum rule for the Wilson coefficients is saturated by the pole at infinity since no leading order cross sections are available.

\subsubsection{ $O^{(1),(8)}_{ud}$}
Just as in the previous section, we obtain (we will omit here flavour indices as these do not play any role, since of the two up quarks and two down quarks should be the same
 to form sum rules)
\bea
&&-2 (c_{ud}^{(1)}-\frac{1}{6}c_{ud}^{(8)}+\frac{1}{2}c_{ud}^{(8)})=\int \frac{ds}{\pi s}\l(\frac{2\sigma^{(8)}_{u \bar d}+\sigma^{(1)}_{u \bar d}}{3}-\sigma _{u  d}^{(6)}\r)+ \frac{1}{3}(2C_\infty^{u\bar{d} (8)}+C_\infty^{u\bar{d} (1)})\nn
&&-2 (c_{ud}^{(1)}-\frac{1}{6}c_{ud}^{(8)})=\int \frac{ds}{\pi s}\l(\sigma _{u \bar d}^{(8)}-\frac{\sigma _{u  d}^{(\bar 3)}+\sigma _{u  d}^{(6)}}{2}\r)+ C_\infty^{u \bar d (8)}
\eea
Rewriting the result in terms of uncolored cross section, we will obtain
\bea
\label{eq:ud}
-2c_{ud}^{(1)}=\int \frac{ds}{\pi s}\l(\sigma_{u \bar d}-\sigma_{u  d}\r)+ \frac{8}{9}C_\infty^{u \bar d (8)}+\frac{1}{9}C_\infty^{u \bar d (1)}
\eea
Interestingly, we see that no constraints can be obtained for $c_{ud}^{(8)}$ if we don't have precise information about the color structure of the initial state. Experiments which are sensitive only to the total scattering cross section will be blind to $c_{ud}^{(8)}$. 

\subsubsection{$O_{eu}, O_{ed}$}
The only operators with sum rule are of the form 
\begin{equation}
(\bar{e}_{Ri}\gamma^\mu e_{Ri})(\bar{u}_{Rja}\gamma_\mu u_{Rja}),
\end{equation}
where no summation over $i,j$ is assumed.
The sum rule is  identical one for both $u$ and $d$ quarks and is given by:
\begin{equation}
-2c^{iijj}_{eu}=\int\frac{ds}{\pi s}\bigg(\sigma_{\bar{e}_{i}u_{j}}-\sigma_{\bar{e}_{i}\bar{u}_{j}}\bigg)+C_\infty^{\bar e u}
\hspace{3mm}\textrm{and}\hspace{3mm}u\leftrightarrow d .
\end{equation}
UV completions are as before, with a positive sign for $u$($d$) coming from a charge $1/3$($4/3$) scalar and a negative sign from a charge $5/3$($2/3$) vector field $V$, note that the amplitude is convergent in the forward limit and the infinite integrals do vanish). Neutral charge $Z'$ can lead to the arbitrary sign of the Wilson coefficient; again, in this case the dispersion relations are saturated by the integrals at infinity.

\subsection{SUM RULES FOR EW DOUBLETS}

In the next 2 subsections we study operators that contribute to doublet-singlet scattering.

\subsubsection{$O_{le},O_{lu},O_{ld},O_{qe}$}
Let us start with the fully leptonic operator and study the forward scattering of $l^pq$ where $p=1,2$ is the isospin, in which case the sum rules are of the form
\bea
  &&  -2c_{le}^{iijj}=\int \frac{ds}{\pi s}\bigg(\sigma_{{l}^{ip}_L \bar e^j_R}-\sigma_{{l}^{ip}_L{e}^j_R}\bigg)+ C_\infty^{l_i e_j}\nn
&&    =\int \frac{ds}{\pi s}\bigg(
\sigma_{e_L^i \bar e^j_R}-
\sigma_{e^{i}_L e^j_R}\bigg)+ C_\infty^{l_i e_j}\nn
&& =\int \frac{ds}{\pi s}\bigg(
\sigma_{\nu_L^i \bar e^j_R}-
\sigma_{\nu^{i}_L e^j_R}\bigg)+ C_\infty^{l_i e_j}
\eea
Similarly, we can write down the sum rules for the the quark lepton operators-
\bea
&&-2c^{iijj}_{lu}=\int\frac{ds}{\pi s}\bigg(\sigma_{\bar{l}^p_i u_{j}}-\sigma_{{l}^p_i{u}_{j}}\bigg)+ C_\infty^{l_i u_j}\hspace{3mm}\textrm{and}\hspace{3mm}u\leftrightarrow d\nn
&&-2c^{iijj}_{qe}=\int\frac{ds}{\pi s}\bigg(\sigma_{\bar{q}^p_{i} e_j}-\sigma_{q^p_{i}e_j}\bigg)+ C_\infty^{q_ie_j},
\eea
where again $p$ stand for the $SU(2)_L$ index. Note that  these sum rules hold true for any isospin for the lepton and any color of the quark.

\subsubsection{$O_{qu}^{(1),(8)}, O_{qd}^{(1),(8)}$}
In this case, the discussion follows closely the one for the quark singlets, and so we arrive at two sum rules(we again suppress the flavour index for brevity) 
\bea
\label{eq:qu}
-2 (c_{qd(u)}^{(1)}-\frac{1}{6}c_{qd(u)}^{(8)}+\frac{1}{2}c_{qd(u)}^{(8)})=\int \frac{ds}{\pi s}\l(\frac{2\sigma^{(8)}_{q \bar d(\bar u)}+\sigma^{(1)}_{q \bar d(\bar u)}}{3}-\sigma _{q d(u)  }^{(6)}\r)+ \frac{1}{3}(2C_\infty^{q\bar{d}(\bar u) (8)}+C_\infty^{q\bar{d}(\bar u) (1)})\nn
-2 (c_{qd(u)}^{(1)}-\frac{1}{6}c_{qd(u)}^{(8)})=\int \frac{ds}{\pi s}\l(\sigma _{q \bar d(\bar u )}^{(8)}-\frac{\sigma_{q  d( u)}^{(\bar 3)}+\sigma_{q d(u)  }^{(6)}}{2}\r)+ C_\infty^{q \bar d (\bar u) (8)}.
\eea
Note that $\sigma_q$ stands for $\sigma_{q^p}$ where $p$
is a $SU(2)$ index and cross sections on the right hand side of the Eq.\ref{eq:qu} can be taken for any component of the quark doublet.
Rewriting the result in terms of uncolored cross section, we will obtain
\bea
\label{eq:qd1}
-2c_{qd(u)}^{(1)}=\int \frac{ds}{\pi s}\l(\sigma_{q \bar d(\bar u )}-\sigma_{q  d (u)}\r)+ \frac{8}{9}C_\infty^{q \bar d (\bar u)(8)}+\frac{1}{9}C_\infty^{q \bar d(\bar u) (1)}.
\eea

Finally, we now study the left handed operators that contribute to doublet-doublet scattering, where the doublet is that of weak isospin.

\subsubsection{$O_{ll}$}
Let us start with the four lepton operator $O_{ll}^{(iijj, ijji)}$. Expanding in components, the following sum rules can be derived (we assume $i\neq j$ and we do not write the operators obtained by interchange of $i\leftrightarrow j$ which are identical, just as in the discussion for up quarks; see Eq. \ref{eq:uu})
\bea
&&-2c^{iijj}_{ll}-2c^{ijji}_{ll}=\int\frac{ds}{\pi s}\bigg(\sigma_{\bar{e_i}e_j}-\sigma_{e_i e_j}\bigg)=\int\frac{ds}{\pi s}\bigg(\sigma_{\bar{\nu}_i\nu_j}-\sigma_{{\nu}_i\bar{\nu}_j}\bigg)+C^{ee,e\nu}_\infty
\nn
&&-2c^{iijj}_{ll}=\int\frac{ds}{\pi s}\bigg(\sigma_{\bar{e_i}\nu_j}-\sigma_{e_i\nu_j}\bigg)+C_\infty^{e\nu}.
\eea
We can decompose the amplitude into the weak isospin amplitudes (see appendix \ref{sec:WE} for details) to obtain the following  dispersion relations
\bea
&&-2c^{iijj}_{ll}-2c^{ijji}_{ll}=\int \frac{d s}{\pi s}\left[\frac{1}{2}\left(\sigma_{i \bar j}^{(1)}+\sigma_{i \bar j}^{(3)}\right)-\sigma_{ij}^{(3)}\right]- C_\infty^{ij (3)}\nn
&&-2c^{iijj}_{ll}=\int \frac{d s}{\pi s}\left[\sigma_{i \bar{j}}^{(3)}-\frac{1}{2}\left(\sigma_{ij}^{(3)}+\sigma_{ij}^{(1)}\right)\right]+C_\infty^{(i \bar j (3))}
\eea
where $(i, j)$ and $(i, \bar j)$ refer to the leptons from $l_i,l_j(\bar l_j)$ doublets and $\sigma^{(3,1)}_{ij,(i \bar j)}$ refers to cross section from the triplet and singlet initial state formed by $ij$ or $i\bar j$. In the case of an operator formed by just one lepton family, we will obtain:
\bea
-8 c_{ll}=\int \frac{d s}{\pi s}\l[\sigma_{e \bar e(\nu \bar \nu)}-\sigma_{ee,(\nu\nu)}\r]+C_\infty^{ee}=\int \frac{d s}{\pi s}\left[\frac{1}{2}\left(\sigma_{l \bar l}^{(1)}+\sigma_{l \bar l}^{(3)}\right)-\sigma_{ll}^{(3)}\right]- C_\infty^{ll (3)}
\nn
-4 c_{ll}=\int \frac{d s}{\pi s}\l[\sigma_{e \bar \nu}-\sigma_{e\nu}\r]+C_\infty^{e\nu}=
\int \frac{d s}{\pi s}\left[\sigma_{l\bar l}^{(3)}-\frac{1}{2}\left(\sigma_{ll}^{(3)}+\sigma_{ll}^{(1)}\right)\right]+C_\infty^{(l \bar l (3))}
\eea

\subsubsection{$O^{(3),(1)}_{lq}$}
In this case, only the operators with $iijj$ flavour structure can contribute; and we arrive at the following dispersion relations-
\bea
&&-2c_{lq}^{(1)}-2c_{lq}^{(3)}=\int\frac{ds }{\pi s}
\l[\sigma_{e \bar d (\nu \bar u)}-\sigma_{e  d (\nu  u)}\r]+ C_\infty^{e \bar d (\nu \bar u)}\nn
&&-2c_{lq}^{(1)}+2c_{lq}^{(3)}=\int\frac{ds }{\pi s}
\l[\sigma_{e \bar u (\nu \bar d)}-\sigma_{e  u (\nu  d)}\r]+ C_\infty^{e \bar u (\nu \bar d)}
\eea
As before, decomposing cross section under isospin we will obtain
\bea
&&-2c_{lq}^{(1)}-2c_{lq}^{(3)}=\int\frac{ds }{\pi s}\l[\frac{1}{2}\l(\sigma_{l\bar q}^{(1)}+\sigma_{l\bar q}^{(3)}\r)-\sigma^{(3)}_{lq}\r]-C_\infty^{lq(3)}\nn
&&-2c_{lq}^{(1)}+2c_{lq}^{(3)}=\int\frac{ds }{\pi s}
\l[\sigma_{l \bar q}^{(3)}-\frac{1}{2}\l(\sigma_{ql}^{(1)}+\sigma_{ql}^{(3)}\r)\r]+ C_\infty^{l\bar q(3)}
\eea

\subsubsection{ $O_{qq}$ and $O^{(3)}_{qq}$}
Let us start with one family,
 in terms of the octet and singlet cross sections,
\bea
-8\l(c_{qq}^{(1)}+c_{qq}^{(3)}\r)=\int\frac{d s }{\pi s}\l[\frac{2\sigma^{(8)}_{u \bar u}+\sigma^{(1)}_{u \bar u}}{3}-\sigma_{uu}^{(6)}\r]-C_\infty^{(6)uu}\nn
-4\l(c_{qq}^{(1)}+c_{qq}^{(3)}\r)=\int\frac{d s }{\pi s}\l[\sigma^{(8)}_{u\bar u}-\frac{\sigma^{\bar 3}_{u u}+\sigma^{(6)}_{uu}}{2}\r]+C_\infty^{(8)u\bar u}
\nn
-4(c_{qq}^{(1)}+c_{qq}^{(3)})=\int\frac{ds}{\pi s}\l[\frac{2\sigma^{(8)}_{u \bar d}+\sigma^{(1)}_{u \bar d}}{3}-\sigma_{ud}^{(6)}\r]-C_\infty^{ud(6)}\nn
-4(c_{qq}^{(1)}-c_{qq}^{(3)})=\int\frac{ds}{\pi s}\l[\sigma^{(8)}_{u \bar d}-\frac{\sigma_{ud}^{(6)}+\sigma_{ud}^{(\bar 3)}}{2}\r]+C_\infty^{ud(8)}
\eea
We can proceed further by performing the double decomposition  in terms of the  $SU(2)_L$ multiplets using the relations
\bea
&&\sigma_{u\bar u}=\frac{1}{2}\l(\sigma^{(1)}_{q\bar q}+\sigma^{(3)}_{q\bar q}\r),~~~\sigma_{u\bar d}=\sigma^{(3)}_{q\bar q}\nn
&& \sigma_{uu}=\sigma^{(3)}_{qq},~~~\sigma_{ud}=\frac{1}{2}\l(\sigma_{qq}^{(1)}+\sigma_{qq}^{(3)}\r).
\eea
Then we will obtain
(the first index will refer now to QCD multiplet and the second one to electroweak).
\bea
-8\l(c_{qq}^{(1)}+c_{qq}^{(3)}\r)=\int\frac{ds}{\pi s}\bigg[\frac{1}{6}\big((2\sigma^{(8,1)}_{q\bar{q}}+\sigma^{(1,1)}_{q\bar{q}}+2\sigma^{(8,3)}_{q\bar{q}}+\sigma^{(1,3)}_{q\bar{q}})\big)-\sigma^{(6,3)}_{qq}\bigg]-C^{(6,3)}_{qq\infty}\nn
-4\l(c_{qq}^{(1)}+c_{qq}^{(3)}\r)=
\int\frac{ds}{\pi s}\bigg[\frac{1}{2}\big(\sigma^{(8,1)}_{q\bar{q}}+\sigma^{(8,3)}_{q\bar{q}}\big)-\frac{1}{2}\big(\sigma^{(\bar{3},3)}_{qq}+\sigma^{(6,3)}_{qq}\big)\bigg]+\frac{C_{q\bar q\infty}^{(8,1)}+C_{q\bar q\infty}^{(8,3)}}{2}
\nn
-4(c_{qq}^{(1)}+c_{qq}^{(3)})=\int\frac{ds}{\pi s}\bigg[\frac{1}{3}\big(2\sigma^{(8,3)}_{q\bar{q}}+\sigma^{(1,3)}_{q\bar{q}}\big)-\frac{1}{2}\big(\sigma^{(6,1)}_{qq}+\sigma^{(6,3)}_{qq}\big)\bigg]-\frac{C_{q q\infty}^{(6,1)}+C_{q q\infty}^{(6,3)}}{2}\nn
-4(c_{qq}^{(1)}-c_{qq}^{(3)})=\int\frac{ds}{\pi s}\bigg[\sigma^{(8,3)}_{q \bar{q}}-\frac{1}{4}\big(\sigma^{(\bar{3},1)}_{qq}+\sigma^{(6,1)}_{qq}+\sigma^{(\bar{3},3)}_{qq}+\sigma^{(6,3)}_{qq}\big)\bigg]+C_{q \bar q\infty}^{(8,3)}
\eea
In terms of the color averaged cross sections,

\bea
\frac{16}{3}\l(c_{qq}^{(1)}+c_{qq}^{(3)}\r)=\int \frac{d s}{\pi s}\l(\frac{\sigma_{q\bar q}^{(3)}+\sigma_{q\bar q}^{(1)}}{2}-\sigma_{qq}^{(3)}\r)-\frac{C_{qq\infty}^{(6,3)}}{3}+\frac{C_{\bar q q\infty}^{(8,1)}+C_{\bar q q\infty}^{(8,3)}}{3}
\nn
-4\l(c_{qq}^{(1)}-\frac{c_{qq}^{(3)}}{3}\r)=\int \frac{d s}{\pi s}\l(\sigma^{(3)}_{q\bar q}-\frac{\sigma_{qq}^{(1)}+\sigma_{qq}^{(3)}}{2}\r)-\frac{C_{qq\infty}^{(6,1)}+C_{qq\infty}^{(6,3)}}{6}+\frac{2C_{q\bar q\infty}^{(8,3)}}{3}
\eea
In the case of two flavours, the disperion relations become:

\bea
-2(c^{iijj}_{qq}+c^{ijji}_{qq}+c^{(3)iijj}_{qq}+c^{(3)ijji}_{qq})=\int\frac{ds}{\pi s}\bigg[\frac{1}{6}\big(2\sigma^{(8,1)}_{q\bar{q}}+\sigma^{(1,1)}_{q\bar{q}}+2\sigma^{(8,3)}_{q\bar{q}}+\sigma^{(1,3)}_{q\bar{q}}\big)-\sigma^{(6,3)}_{qq}\bigg]-C_{qq\infty}^{(6,3)}
\nn
-2(c^{iijj}_{qq}+c^{(3)iijj}_{qq})=\int\frac{ds}{\pi s}\bigg[\frac{1}{2}\big(\sigma^{(8,1)}_{q\bar{q}}+\sigma^{(8,3)}_{q\bar{q}}\big)-\frac{1}{2}\big(\sigma^{(\bar{3},3)}_{qq}+\sigma^{(6,3)}_{qq}\big)\bigg]+\frac{C_{ q \bar q\infty}^{(8,1)}+C_{q\bar q\infty}^{(8,3)}}{2}
\nn
-2(c^{iijj}_{qq}-c^{(3)iijj}_{qq}+2c^{(3)ijji}_{qq})=\int\frac{ds}{\pi s}\bigg[\frac{1}{3}\big(2\sigma^{(8,3)}_{q\bar{q}}+\sigma^{(1,3)}_{q\bar{q}}\big)-\frac{1}{2}\big(\sigma^{(6,1)}_{qq}+\sigma^{(6,3)}_{qq}\big)\bigg]-\frac{C_{qq\infty}^{(6,1)}+C_{qq\infty}^{(6,3)}}{2}
\nn
-2(c^{iijj}_{qq}-c^{(3)iijj}_{qq})=\int\frac{ds}{\pi s}\bigg[\sigma^{(8,3)}_{q \bar{q}}-\frac{1}{4}\big(\sigma^{(\bar{3},1)}_{qq}+\sigma^{(6,1)}_{qq}+\sigma^{(\bar{3},3)}_{qq}+\sigma^{(6,3)}_{qq }\big)\bigg]-C_{q \bar q\infty}^{(8,3)}\nn
 \eea
The power of these relations
relations allows to understand immediately the signs of the Wilson coefficients  in the various UV completions. 
For example, for a scalar diquark which is in $(\bar 6,1,-1/3)$ representation under $SU(3)\times SU(2)\times U(1)_Y$
we will get:
\bea
c^{iijj}_{qq, \bar6}=c^{(3)ijji}_{qq,\bar 6}=-
c^{(3)iijj}_{qq,\bar 6}=-c_{qq,\bar 6}^{ijji}>0.
\eea
Similarly, for 
a scalar diquark which is in $( 3,1,-1/3)$  will get:
\bea
c^{iijj}_{qq, 3}=c_{qq, 3}^{ijji}=-c^{(3)ijji}_{qq, 3}=-
c^{(3)iijj}_{qq, 3}>0.
\eea
Finally, we can sum and report these sum rules in terms of color averaged cross sections, which yield 2 equations depending on whether the initial and final state form $SU(2)_L$ triplets or singlets.
\bea
\label{eq:qq}
-2(c^{iijj}_{qq}+c^{(3)iijj}_{qq}+\frac{1}{3}c^{ijji}_{qq}+\frac{1}{3}c^{(3)ijji}_{qq})=\int \frac{d s}{\pi s}\l[\frac{\sigma_{q\bar q}^{(3)}+\sigma_{q\bar q}^{(1)}}{2}-\sigma_{qq}^{(3)}\r]-\frac{C_{qq}^{(6,3)}}{3}+\frac{C_{\bar q q\infty}^{(8,1)}+C_{\bar q q\infty}^{(8,3)}}{3},
\nn 
-2(c^{iijj}_{qq}-c^{(3)iijj}_{qq}+\frac{2}{3}c^{(3)ijji}_{qq})=\int \frac{d s}{\pi s}\l[\sigma^{(3)}_{q\bar q}-\frac{\sigma_{qq}^{(1)}+\sigma_{qq}^{(3)}}{2}\r]-\frac{C_{qq\infty}^{(6,1)}+C_{qq\infty}^{(6,3)}}{6}+\frac{2C_{q\bar q\infty}^{(8,3)}}{3}.\nonumber\\
\eea

\section{Summary}
\label{sec:sum}

In this work, we explored the sum rules for four-fermion operators at dimension six level.
As expected, the  convergence of the dispersion integrals leading to the dimension six 
Wilson coefficients is not guaranteed, and in particular is spoiled by the t-channel 
exchange of the vector bosons. This additional feature can 
modify the predictions of the dispersion relations for sign and strength of IR interactions, and 
for some UV completions the value of the Wilson 
coefficients can be even  saturated by the pole at infinity.
However we find that this ambiguity of 
IR couplings is not related 
to the (non)convergence  of the dispersion integrals  and 
as an example, we have constructed, in addition to tree level, 1-loop  weakly coupled models  (see section  \ref{sec:loop}) 
where both signs become available even when the integral 
over the infinite circle vanishes.

We presented forward dispersion relations for all possible four-fermion dimension six operators. To facilitate the connection between the values of the Wilson coefficients and new physics scenarios,  we have 
performed the decomposition in terms of the $SU(2)$ and $SU(3)$ multiplets. 
Such relations predict in a model independent way processes with enhanced cross section in the case of discoveries in low energy experiments.
We carefully indicate all the relevant quantum numbers of the quantities 
involved in our dispersion relations in order to provide a convenient dictionary for future measurements, where the 
precise structure of initial states is often unavailable. This can have interesting consequences; for example, 
Eq.\ref{eq:ud} tells us that measuring uncoloured cross sections in the UV clouds any information about 
$c^{(8)}_{ud,(qu),(qd)}$ Wilson coefficients, despite it contributing formally to sum rules with  fixed initial colours.

We emphasize 
that these sum rules are to be interpreted as a model independent link between UV and IR measurements,
as opposed to the usual positivity bounds. Even though less constraining on the EFT parameter space, these relations  can instead be
 used as a powerful tool for model building to unearth 
the underlying, fundamental physics that is to be explored in the coming years.

\section*{Acknowledgements}
AA in part was supported by the MIUR contract 2017L5W2PT. 
DG acknowledges support through the Ramanujan Fellowship and MATRICS Grant of the Department of Science and Technology, Government of India.
We would like to thank Joan Elias Miro for discussion and comments.

\appendix
\section{Massless spinor helicity conventions}
\label{sec:aphelicity}
We will briefly summarize the key results relevant to us (for a pedagogical introduction see \cite{Elvang:2013cua} ) in the $(+,-,-,-)$ signature (we will follow the conventions discussed in \cite{Baratella:2020lzz, Dreiner:2008tw,Mangano:1990by}). We have the 2 component spinors $v_{L/R}, u_{L/R}$ and their barred versions. They are related by crossing symmetry, $u_{L/R}=v_{R/L}, \bar{u}_{L/R}=\bar{v}_{R/L}$. It is important to realise that for antiparticles, the spinor has opposite handedness to the field that describes it. For instance, a right chiral field $e_R$ has an antiparticle which has the spinor $v_L$, while the particle carries the spinor $u_R$. In other words, both $u_R, v_L$ correspond to a right chiral field; whereas $v_R, u_L$ correspond to a left chiral field. To be absolutely clear, we will just refer to the handedness of the relevant spinor as opposed to the helicity of a particle/antiparticle wherever necessary.

Operationally, we will assign the brackets 
\begin{equation}
    \bar{v}_L=\bar{u}_R\equiv [,\hspace{3mm}\bar{v}_R=\bar{u}_L\equiv\langle, \hspace{3mm}v_L=u_R\equiv\rangle,\hspace{3mm}v_R= u_L\equiv ].
\end{equation}

The inner product is antisymmetric-as is expected for grassman-valued quantities-

\begin{equation}
    \langle pq\rangle=-\langle qp\rangle \hspace{5mm}[pq]=-[qp]
\end{equation}

Note that this also means that $\langle pp\rangle=0=[pp]$. Mixed brackets vanish. The formalism encodes a lot of power-for example, it tells us that a $\langle$ and $]$ type spinor cannot occur at a vertex unless there's a $\gamma^\mu$ involved-a vector connects opposite helicity particles. Similarly, same helicity spinors making up a vertex indicate a scalar is involved. 

We will not insist on taking all momenta ingoing/outgoing; in our calculations, the momenta labelled $1,2$ are always incoming and $3,4$ are always outgoing. We can freely work with negative momenta via the standard analytic continuation-

\begin{equation}
    |-p\rangle=i|p\rangle\hspace{5mm}|-p]=i|p]
\end{equation}

These brackets satisfy the property

\begin{equation}
    \langle 1|\gamma^\mu 2]=[2|\gamma^\mu 1\rangle
\end{equation}

Furthermore, we have 
\begin{equation}
[i|\gamma_\mu|i\rangle=2p_i\hspace{5mm}\langle ij\rangle[ij]=-2p_i\cdot p_j=(p_i-p_j)^2
\end{equation}

since $p_i^2=0$ for massless spinors. We therefore have our mandelstam variables-

\begin{equation}
    s=2p_1\cdot p_2=-[12]\langle 12\rangle \hspace{4mm} t=-2p_3\cdot p_1=[13]\langle 13\rangle\hspace{4mm} u=-2p_4\cdot p_1=[14]\langle 14\rangle
\end{equation}

Finally, we have the all important Fierz rearrangement-

\begin{equation}
  [1|\gamma^\mu|2\rangle[3|\gamma_\mu|4\rangle=-2[13]\langle 24\rangle  
\end{equation}

\section{Details about cross sections and loop amplitudes}
\label{sec:brutef}
In this appendix we will give details about explicit verification of the dispersion relations presented in the text for various models.
\subsection{$Z'$ at tree level}
Let us start with   neutral vector $Z'$ coupled to right-handed current via $\lambda Z'_\mu\bar{e}_R\gamma^\mu e_R$. It generates $e_R  \overline{ e_R}$ scattering through the diagrams 
\begin{figure}[htp]
    \centering
    \includegraphics[width=8cm]{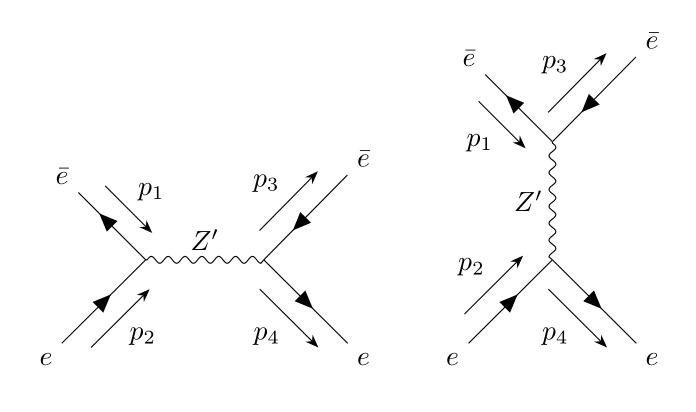}
    \caption{Fermion statistics dictate these diagrams subtract}
    \end{figure}
The full amplitude  will be given by
\begin{equation}
 A_{e\bar e}=-2\lambda^2[14]\langle 23\rangle\bigg(\frac{1}{s-m^2}+\frac{1}{t-m^2}\bigg)
\end{equation}
Matching the IR and UV amplitudes at low energies we will obtain
\begin{equation}
-8c^{1111}_{ee}[14]\langle23\rangle=-2\lambda^2[14]\langle 23\rangle\bigg(\frac{1}{-m^2}+\frac{1}{-m^2}\bigg)\implies c^{1111}_{ee}=-\frac{\lambda^2}{2m^2}
\end{equation}
Let us verify that this is consistent with our dispersion relation. With a vector $Z'$ at order $O(\lambda^2)$ in perturbation theory we have
$\sigma_{e\bar e}\neq 0$ and$\sigma_{ee}=0$.
To calculate the cross sections, note that by the optical theorem, we have 
\begin{equation}
Im(e \bar e\to e\bar e)=s\sigma^{tot}_{e\bar e }
\end{equation}
We use the fact that $Im\bigg(\frac{1}{p^2-m^2+i\epsilon}\bigg)=-\pi\delta(p^2-m^2)$ which, when substituted in the amplitude (\ref{eq:AmpZ}) gives us

\begin{equation}
Im(e^+_Le^-_R\to e^+_L e^-_R)=2\lambda^2\pi s\delta(s-m^2)
\end{equation}
Starting from dispersion relation in Eq.\ref{eq:dispmaster} we will get,
\begin{equation}
-8c^{1111}_{ee}=\int \frac{ds}{\pi s}(\sigma_{e\bar e}-0)+C_\infty=\int \frac{ds}{\pi s^2}Im(e\bar e\to e\bar e)+C_\infty=\frac{2\lambda^2}{m^2}+C_\infty.
\end{equation}
Calculating explicitly $C_\infty$ we will obtain:
\bea
&&C_\infty=\int^{2\pi}_0 \frac{d\theta}{2\pi}\frac{A(|s_\Lambda|e^{i\theta},0)}{(|s_\Lambda|e^{i\theta})^{2}}\cdot(|s_\Lambda|e^{i\theta})=\frac{2\lambda^2}{m^2}\nn
&&A|_{t\to 0}=-2 \lambda^2 s\l(\frac{1}{s-m^2}+\frac{1}{-m^2}\r)
\eea
Which is of the same sign as the dispersion integral, and therefore we find
\begin{equation}
    -8c^{1111}_{ee}=4\lambda^2/m^2\implies c^{1111}_{ee}=-\lambda^2/2m^2
\end{equation}
as claimed in (\ref{eq:crr}), and our dispersion relation is explicitly verified.

\subsection{Integrating out color octet}\label{sec:octet}
Very similarly to the charge neutral $Z'$ we can consider effects coming from integrating out color octet $V$ which has zero electric charge. Let us look for example on octet interacting with right -haned up quark current:
\begin{equation}
    g_{ij}V^A_\mu(\bar{u}_i\gamma^\mu T^A u_j)\implies c^{ijkl}_{uu}=\frac{-g_{kj}g_{il}}{M_V^2}+\frac{g_{ij}g_{kl}}{3M_V^2}.
\end{equation}
Let us assume that the octet couplings are universal and flavour diagonal,  then $g_{ij}=g\delta_{ij}$, and the Wilson  coefficients are equal to 
\begin{equation}
    c^{iijj}_{uu}=\frac{2g^2}{3M_V^2},~~~ c^{ijji}_{uu}=\frac{-2g^2}{M_V^2}
\end{equation}
Now let us look at dispersion relations for $i\neq j$, then similar to the discussion in Eq.\ref{eq:zprime2}, the cross sections will vanish at $O(g^2)$ and the right hand side of Eq. \ref{eq:uu} will be controlled by the contribution of the integrals over infinite contours.
\begin{equation}
  C_\infty^{(8)}  =C_\infty^{\alpha\neq\beta}=\frac{8g^2}{3M^2_V},~~~~ C_\infty^{\alpha\alpha}=-C_{\infty}^{(6)}=-\frac{4g^2}{3M_V^2}
\end{equation}
which confirm the dispersion relations 
\bea
\label{eq:uu}
&&-2 c_u^{iijj}=- C_\infty^{uu,(6)}
\nn
&&-2 (c_u^{iijj}+c_u^{ijji})= C_\infty^{u\bar u,(8)}.
\eea

\subsection{Charge 2 scalar at tree level}
Let us  build a model where only $\sigma_{ee (\bar e \bar e )}$ is present at the lowest order in perturbation theory. This can be done with 
 a charge (2) scalar, which interacts as follows $(\lambda \phi\bar{e}_Re_R^c+h.c)$ where the $c$ subscript stands for charge conjugation. 
\begin{figure}[htp]
    \centering
    \includegraphics[width=4cm]{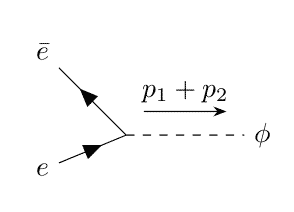}
    \caption{ Scalar production in $\bar e\bar e$ collision. note that the vertex is     $=2!(-i\lambda)$.}
    \end{figure}
Matching  the amplitudes in EFT and UV theory we will obtain
\begin{equation}
-8 c^{1111}_{ee}[14]\langle 23\rangle=-2!2!\lambda^2\frac{[14]\langle 32\rangle}{-m^2}\implies c^{1111}_{ee}=+\frac{\lambda^2}{2m^2}
\end{equation}
Then the scattering cross section is equal to:
 \begin{equation}
\sigma^{tot}_{\bar e\bar e}=4\lambda^2\pi\delta(s-m^2)  .
 \end{equation}
So that dispersion relation becomes:
 \begin{equation}
-8c^{1111}_{ee}=\int \bigg(0-\frac{ds}{\pi s}\sigma_{++}\bigg)=-\frac{4\lambda^2}{m^2}
\end{equation}
and as expected we find:
 $c^{1111}_{ee}=+\frac{\lambda^2}{2m^2}$.

\subsection{Dispersion relation at 1-loop }
At last let us consider the following UV completion for the $(\bar e \gamma_\mu e )(\bar e \gamma_\mu e)$ operator. It will demonstrate that 
 it is possible to have a negative Wilson coefficient
 with vanishing integrals over infinite crircles.
Let us extend SM with a new heavy scalar and fermion with interactions 
\bea
\lambda (\Phi\bar{e}_R\Psi)+h.c,
\eea
where electric charges of new fields satisfy $Q[\Phi]+Q[\Psi]=-1$.
Let us start  by deriving the $c_{ee}$ Wilson coefficient.

We consider $e\bar e\to e\bar e$ scattering; then the amplitude will be given by a box diagram and it's crossed version .
\begin{figure}[htp]
    \centering
    \includegraphics[scale=1]{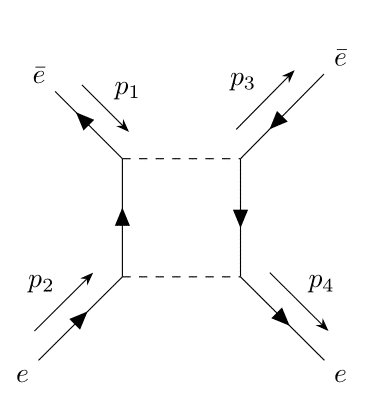}
    \caption{Forward amplitude at $O(\lambda^4)$ order.
    }\label{fig:box}
    \end{figure}
In order to match with EFT predictions, we can focus on the processes where external particles have vanishing momentum, in which case the amplitude will be given by
\begin{equation}
 iM=\lambda^4 [1|\gamma_\mu|2\rangle[4|\gamma_\nu|3\rangle \int \frac{d^D k}{(2\pi)^D}\frac{k^\mu k^\nu}{(k^2-m^2)^4}-(2\leftrightarrow 3).
\end{equation}
Now, we have assumed that the masses of the new fields are equal $m[\Phi]=m[\Psi]=m$; the loop function for arbirary masses is reported in the main text. Performing the integral , which is finite, and doing the Fierz rearrangements we will obtain:
\begin{equation}
M=\frac{1}{3}\frac{\lambda^4}{16\pi^2 m^2}[14]\langle 23\rangle, \Rightarrow c^{1111}_{ee}=-\frac{1}{3}\frac{\lambda^4}{128\pi^2 m^2}
\end{equation}
So we see that sign of the Wilson coefficient is indeed negative.

By looking at the amplitude at $s\to \infty$ we can see that $A(s)/s\to 0$ at infinite circle, so all we need to know is the cross section for $e \bar e$ scattering to verify the dispersion relations.  The total cross section at the order $O(\lambda^4)$ will be given by the two processes $e\bar e\to \Psi \bar \Psi$ and $e \bar e\to \Phi \Phi^*$, and there will be no processes $e e \to$ anything at $O(\lambda^4)$. 
\begin{figure}
    \centering
    \includegraphics[scale=1]{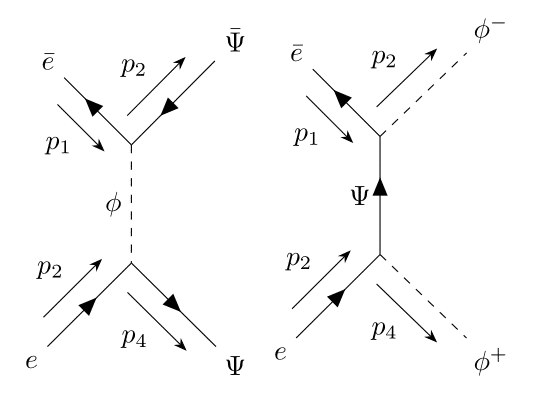}
    \caption{Processes contributing to the $\sigma_{e\bar e}$ at $O(\lambda^4)$ order.}
    \label{fig:my_label}
\end{figure}
Performing the calculation we obtain
\bea
&&\sigma(e \bar e\to \Psi\bar \Psi)=\frac{\lambda^4}{16\pi s^2}\sqrt{(s(s-4m^2)}\nn
&&\sigma (e \bar e\to \Phi \Phi^*)=\frac{\lambda^4}{64\pi s^2}\bigg(-8\sqrt{s(s-4m^2)}-4s\log\bigg(\frac{s-\sqrt{s(s-4m^2)}}{s+\sqrt{s(s-4m^2}}\bigg)\bigg)
\eea
Performing the calculation for  the dispersion integral we will obtain:
\begin{equation}
\int\frac{ds}{\pi s}(\sigma(e \bar e\to \Psi\bar \Psi)+\sigma(e \bar e\to \Phi \Phi^*))=\frac{\lambda^4}{\pi^2m^2}(1/96+1/96)=\frac{\lambda^4}{48m^2\pi^2}=-8 c_{ee}^{1111}
\end{equation}
satisfying the identity of Eq.\ref{eq:dispmaster}.

\section{Decomposition of cross sections in terms of $SU(2)$ and $SU(3)$ irreps}\label{sec:WE}
In this section, we will give details of the 
decomposition of amplitudes in terms of the irreducible representations of the 
electoweak $SU(2)$ and QCD 
$SU(3)$ groups. The Wigner-Eckart theorem tells us 
that the resulting amplitudes and cross 
sections will depend only on representations of the 
initial state (see for similar decompositions of the isospin \cite{Olsson:1967vrz,PhysRev.140.B736}, custodial \cite{Falkowski:2012vh,Urbano:2013aoa} and other groups \cite{Trott:2020ebl,Bellazzini:2014waa}).
Let us start with two lepton doublet scattering
$L_{1} L_{2} \rightarrow L_{1} L_{2}$ where $L_{1}, L_{2}$ are $S U(2)_{L}$ doublet leptons, for eg $(\nu_e, e)^T$. Then, the initial state can be decomposed as a singlet and a triplet under SU(2): $2\otimes 2=3\oplus 1$
where the singlet and triplet states are defined as follows:
\bea
S=\textrm{singlet}=\frac{1}{\sqrt{2}}(|\nu e\rangle-|e\nu\rangle) \nn
T=\textrm{ triplet }=\left\{\begin{array}{c}
|\nu \nu\rangle \\
\frac{1}{\sqrt{2}}(\left|\nu e\rangle+|e \nu\right\rangle) \\
|e e\rangle
\end{array}\right., 
\eea
where $(\nu,e)$ are the components of EW doublet. Similarly, we can decompose the states for the lepton and anti-lepton scattering, where we will find:
\bea
L_{1}=(\nu_{1}, e_{1})^{T}, \hspace{5mm} \bar{L}_{2}=(-\bar{e}_{2},\bar{\nu}_{2})^{T} \\
\tilde{S}=\textrm{singlet}=\frac{1}{\sqrt{2}}(|e \bar{e}\rangle+|\nu \bar{\nu}\rangle) \\
\tilde{T}=\text { triplet }=\left\{\begin{array}{c}
-|\nu \bar{e}\rangle \\
\frac{1}{\sqrt{2}}(|\nu \bar{\nu}\rangle-|e \bar{e}\rangle) \\
|e \bar{\nu}\rangle
\end{array}\right.
\eea
Using this decomposition we can immediately see that the amplitude for the forward scatterings of the various components of the doublets will be decomposed as
\bea
A_{e e}=A_{LL}^{(3)}, \quad A_{e \bar{e}}=\frac{A_{L\bar L }^{(3)}+A_{L \bar{L}}^{(1)}}{2},\nn
A_{\nu e}=\frac{A_{LL}^{(1)}+A_{LL}^{(3)}}{2},~~A_{\bar \nu  e}=A_{L\bar L}^{(3)},
\nn
A_{\nu \nu}=A_{LL}^{(3)}, \quad A_{\nu \bar{\nu}}=\frac{A_{L\bar L }^{(3)}+A_{L \bar{L}}^{(1)}}{2}
\eea
and similarly, we can decompose the cross sections for quark lepton doublet scatterings.
Note that forward amplitudes  will satisfy the following crossing relations:
\bea
A_{LL}^{(3)}(s,u)=
\frac{A_{L\bar L}^{(3)}(u,s)+A_{L\bar L}^{(1)}(u,s)}{2}
,~~~\frac{A_{L L}^{(3)}(s,u)+A_{L L}^{(1)}(s,u)}{2}=A_{L\bar L}^{(3)}(u,s).
\eea
Since  we are looking at the dispersion relations for dimension six operators and the amplitudes in IR scale linearly with $s$, the integrals over infinite circle contours must satisfy:
\bea
&&C_\infty^{LL(L\bar L)}\equiv \int_{\rm infinite ~circle}\frac{ds }{s^2} A^{LL(L\bar L)}(s),\nonumber\\
&&-C_\infty^{LL(3)}=
\frac{ C_\infty^{L\bar L (3)}+C_\infty^{L\bar L (1)}}{2},~~~ -\frac{ C_\infty^{L L (3)}+C_\infty^{L L (1)}}{2}=C_\infty^{L\bar L(3)}.
\eea
The situation is very similar for the quark quark doublet scattering but there we can decompose the initial state in the  representations of the color  $SU(3)$  as well (see \cite{Trott:2020ebl} for an example).

\subsection{$SU(3)$ decomposition}
Let us consider for simplicity scattering of the quarks which are singlets under $SU(2)$, in which case
\bea
3\otimes 3= \bar 3\oplus 6,~~~3\otimes \bar 3 =1\oplus 8
\eea
In the case of two particle scattering, the only two possibilities are when initial particles have the same, or different colors. For the quark antiquark scattering,various initial color states can be decomposed as
\bea
&&|1 \bar 1\rangle=\frac{S}{\sqrt 3}+ \frac{\lambda_8}{\sqrt 6}+\frac{\lambda_2}{\sqrt 2},~~
|2\bar 2\rangle=\frac{S}{\sqrt 3}+ \frac{\lambda_8}{\sqrt 6}-\frac{\lambda_2}{\sqrt 2}\nn &&
|3 \bar 3\rangle=\frac{S-\sqrt 2 \lambda_8}{\sqrt 3},~~|1\bar 2\rangle = \frac{\lambda_1+i \lambda_2}{\sqrt 2},~~|2\bar 1\rangle = \frac{\lambda_1-i \lambda_2}{\sqrt 2}\nn &&
|1\bar 3\rangle= \frac{\lambda_4+i \lambda_5}{\sqrt 2},~~|3\bar 1\rangle= \frac{\lambda_4-i \lambda_5}{\sqrt 2},~~|2\bar 3\rangle= \frac{\lambda_6+i \lambda_7}{\sqrt 2},~~|3\bar 2\rangle= \frac{\lambda_6-i \lambda_7}{\sqrt 2}
\eea
Where
$
S=\frac{|1\bar1 \rangle+ |2\bar2\rangle+|3\bar3\rangle}{\sqrt 3}
$
 is a $SU(3)$ singlet state  and $(\lambda_1...\lambda_8)$ are components of an octet, which can be formed Using Gell Mann matrices (our normalization is $\langle\lambda_i|\lambda_j\rangle=\delta_{ij})$. Similarly,we can decompose the quark-quark initial state in terms of the $\bf 6$ and $\bf \bar 3$  . Note that in this case, the same and different color initial states can be schematically decomposed as 
 \bea
 |\alpha \alpha \rangle  = {\bf 6}_{\alpha \alpha}, |\alpha \beta\rangle_{\alpha\neq \beta}=\frac{{\bf 6}_{\alpha \beta}\pm \bar{\bf 3}_{\alpha \beta}}{\sqrt 2}
 \eea
 Then, the Wigner Eckart theorem tells us that the total cross sections and forward scattering amplitudes will satisfy the following relations:
\begin{align}
    \sigma_{\alpha \alpha}=\sigma^{(6)},~~
    \sigma_{\alpha \beta}|_{\alpha \neq \beta}=\frac{1}{2}(\sigma^{(\bar{3})}+\sigma^{(6)}), \\
    \sigma_{\alpha\bar{\alpha}}=\frac{\sigma^{(1)}_{}+2\sigma^{(8)}}{3},~~~
    \sigma_{\alpha\bar{\beta}}|_{\alpha\neq \beta}=\sigma^{(8)},
\end{align}
where  ${\alpha (\bar \beta)}$ indices indicate whether we are looking at the same or different color scatterings in $qq$, or $q\bar q$ channels ($q$ here stands for a quark, which can be either up or down type).  In case we are interested in the color averaged cross sections, these will be related to the above as follows
\bea
\sigma_{qq}\equiv \l(\sigma_{qq}\r)_{col.aver.}=\frac{2}{3}\sigma^{(6)}+\frac{1}{3}\sigma^{(\bar 3)}\nn
\sigma_{q\bar q}\equiv \l(\sigma_{q\bar q}\r)_{col.aver.}=\frac{1}{9}\sigma^{(1)}+\frac{8}{9}\sigma^{(8)}\nn
\eea
At last, forward amplitudes decomposed under QCD representations will satisfy the following crossing relations:
\bea
A_{qq}^{(6)}(s,u)=\frac{A_{q\bar q}^{(1)}(u,s)+2 A_{q\bar q}^{(8)}(u,s)}{3},~~~\frac{A_{qq}^{(\bar 3)}(s,u)+A_{qq}^{(6)}(s,u)}{2}=A_{q\bar q}^{(8)}(u,s)
\eea
Similarly, the contours over the infinite circles will be related as follows:
\bea
-C_{\infty}^{qq(6)}=\frac{C_{\infty}^{q\bar q(1)}+2 C_{\infty}^{q\bar q(8)}}{3},~~~-\frac{C_{\infty}^{qq(\bar 3)}+C_{\infty}^{qq(6)}}{2}=C_{\infty}^{q\bar q(8)}
.
\eea

\color{black}
\providecommand{\href}[2]{#2}\begingroup\raggedright\endgroup

\end{document}